\documentclass[fleqn,twoside]{article}
\usepackage{espcrc2}
\usepackage{epsf}

\newcommand{\BE}{\begin{equation}}
\def\EE{\end{equation}}
\def\BEA{\begin{eqnarray}}
\def\EEA{\end{eqnarray}}
\def\EL{\nonumber\\}

\newsavebox{\Staple}
\savebox{\Staple}{\begin{picture}(0,0)
\thicklines
\put(0.0,0.1){\vector(0,1){0.9}}
\put(0.0,1.0){\vector(1,0){0.9}}
\put(0.9,1.0){\vector(0,-1){0.9}}
\end{picture}}

\newsavebox{\InvertedStaple}
\savebox{\InvertedStaple}{\begin{picture}(0,0)
\thicklines
\put(0.0,-0.1){\vector(0,-1){0.9}}
\put(0.0,-1.0){\vector(1,0){0.9}}
\put(0.9,-1.0){\vector(0,+1){0.9}}
\end{picture}}

\newsavebox{\FiveStaple}
\savebox{\FiveStaple}{\begin{picture}(0,0)
\thicklines
\put(0.0,0.1){\vector(0,1){0.9}}
\put(0.0,1.0){\vector(1,1){0.5}}
\put(0.5,1.5){\vector(1,0){0.9}}
\put(1.4,1.5){\vector(-1,-1){0.5}}
\put(0.9,1.0){\vector(0,-1){0.9}}
\end{picture}}

\newsavebox{\SevenStaple}
\savebox{\SevenStaple}{\begin{picture}(0,0)
\thicklines
\put(0.0,0.1){\vector(0,1){0.9}}
\put(0.0,1.0){\vector(1,1){0.5}}
\put(0.5,1.5){\vector(1,2){0.3}}
\put(0.8,2.1){\vector(1,0){0.9}}
\put(1.7,2.1){\vector(-1,-2){0.3}}
\put(1.4,1.5){\vector(-1,-1){0.5}}
\put(0.9,1.0){\vector(0,-1){0.9}}
\end{picture}}

\newsavebox{\LepageStaple}
\savebox{\LepageStaple}{\begin{picture}(0,0)
\thicklines
\put(0.0,0.1){\vector(0,1){0.9}}
\put(0.0,1.0){\vector(0,1){0.9}}
\put(0.0,1.9){\vector(1,0){0.9}}
\put(0.9,1.9){\vector(0,-1){0.9}}
\put(0.9,1.0){\vector(0,-1){0.9}}
\end{picture}}

\newsavebox{\Link}
\savebox{\Link}{\begin{picture}(0,0)
\thicklines
\put(0.0,0.0){\vector(1,0){0.9}}
\end{picture}}

\newsavebox{\Naik}
\savebox{\Naik}{\begin{picture}(0,0)
\thicklines
\put(0.0,0.0){\vector(1,0){0.9}}
\put(1.0,0.0){\vector(1,0){0.9}}
\put(2.0,0.0){\vector(1,0){0.9}}
\end{picture}}

\title{Results on improved KS dynamical configurations: spectrum, 
decay constants, etc.}
\author{Steven Gottlieb 
\address{Department of Physics, Indiana University, Bloomington, IN 47405, USA}
}
\begin{document}

\begin{abstract}
The MILC Collaboration has been producing ensembles of lattice
configurations with three dynamical flavors for the past few years.
There are now results for three lattice spacings for a variety of
light and strange quark masses, ranging down to $m_l=0.1 m_s$, where
$m_s$ is the dynamical strange quark mass and $m_l$ is the common mass
of the $u$ and $d$ quarks.  Recently, the Fermilab, HPQCD, MILC and UKQCD
collaborations have presented a summary of results obtained using these
lattices.  
Compared with quenched results,
these new calculations show great improvement in agreement with experiment.
This talk addresses the technical improvements that make these calculations
possible and provides additional details of calculations not presented
in the initial summary.  We demonstrate that a wide range of hadronic
observables can now be calculated to 2--3\% accuracy.
\end{abstract}

\maketitle

\section{INTRODUCTION}
The past several years have seen a marked improvement in our ability to
do realistic calculations in Lattice QCD (LQCD).  Improved actions have
better scaling properties thus allowing us to more quickly approach the
continuum limit than with the simplest actions.  For staggered quarks,
these improved actions also allow us to perform simulations (on
current computers) with small enough up and down quark masses that we
can use chiral perturbation theory to extrapolate to the physical
quark masses.  Further, the use of improved actions has greatly reduced the
``flavor'' symmetry violations, that we now call ``taste'' symmetry
violations.  Without this improvement, many of the non-Goldstone pion
states are as heavy as the lightest kaon state.  However, with
the improvement in taste symmetry, the splitting between pions and kaons is
more clear, and one expects to be better able to discern the effects of a
dynamical strange quark.  These improvements, and the general increase in
available computer power have made it attractive to embark on a large scale
calculation of QCD with three dynamical quarks.  

Generation of the gauge ensembles has been going on for four years.  MILC
has been sharing dynamical configurations through the Gauge Connection and
informally for some time.  
The Fermilab, HPQCD, MILC and UKQCD collaborations have all been using the
MILC configurations to calculate a variety of quantities with both light
($u$, $d$ or $s$) and heavy ($c$ or $b$) quarks.  Taken together, we find
that these calculations result in a compelling picture.  LQCD with three
flavors agrees with experiment with errors of 2--3\% for a wide variety of
quantities, whereas quenched QCD has errors as large as 15--20\%.
%%%FIRST TO CUT
%%%Although there is much work to be done to calculate other observables,
%%%many of which are intrinsicly more difficult to accurately calculate on
%%%the lattice, I was asked to review these developments with staggered quarks.
%%%I offer my
%%%apologies to those who would prefer a review of results with all approaches
%%%to dynamical quarks, and to those other groups who are also using MILC
%%%ensembles for a variety of interesting studies.  As interesting as those
%%%topics are, they are not what the Local Organizing Committee asked me to
%%%discuss.

In the rest of this article, I shall briefly discuss some of the issues
that arise when dealing with staggered quarks.  I will then give an
introduction to the parameters set in lattice calculations, and some
details of the MILC ensembles.  After an explanation of high precision
observables, we go on to discuss a variety of calculations such as the
light meson decay constants, the heavy quark spectrum, the light quark spectrum
and topological susceptibility.  We conclude with a discussion of future
prospects.

%\item Charge of LOC
%\item Algorithm Issues
%\item Introduction to Lattice Calculations
%\item The Ensemble of Configurations 
%\item High Precision Observables
%\item Light Meson Decay Constants and Chiral Extrapolations
%%\item Exotics ?
%\item Spectrum including Heavy Quarks
%\item Light Quark Spectrum
%%%\item $\alpha_s$
%\item Topology 
%\item{Prospects}

\section{ALGORITHM ISSUES}

Staggered quarks are fast compared to other discretizations, but
one must take a fourth root to deal with the so-called fermion
doubling problem.  This leads to a potential loss of
locality, and we don't know how to construct a transfer matrix.
Nevertheless, there are positive things one can say.

If one considers a formal perturbative expansion of the theory,
the fractional root causes no problems in the expansion.

It is known that the CP violation that should occur 
%%for a theory with an odd number of quarks 
when $m_u+m_d <0$ does not happen
with staggered quarks, but this deficiency is not obviously relevant 
to the real world.

The taste singlet axial current is only partially conserved so the chiral
anomaly is preserved.

We now understand how to modify chiral perturbation theory in the presence
of taste changing interactions  \cite{Bernard:2001yj,Aubin:2002ss}.

An important part of recent progress in LQCD is the use of improved actions.
The MILC collaboration uses the ``Asqtad'' action 
\cite{Blum:1996uf,Lagae:1998pe,Lepage:1998vj,Orginos:1999cr,Bernard:1999xx}.

The gauge action includes the plaquette, the
$1\times 2$ rectangle and a bent parallelogram 6-link term.  The
quark action includes paths up to seven links long as shown below 
and the 3-link Naik term.

% link, staple, 5staple, 7staple
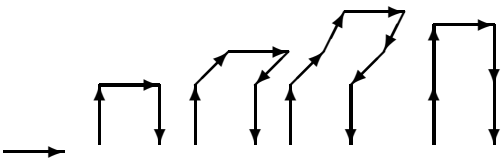
\begin{figure}[b]
\vspace{-30pt}
\setlength{\unitlength}{0.5in}
\begin{center}\begin{picture}(6.0,2.0)
\put(0.0,0){\usebox{\Link}\makebox(0,0)}
\put(1.0,0){\usebox{\Staple}\makebox(0,0)}
\put(2.0,0){\usebox{\FiveStaple}\makebox(0,0)}
\put(3.0,0){\usebox{\SevenStaple}\makebox(0,0)}
\put(4.5,0){\usebox{\LepageStaple}\makebox(0,0)}
\end{picture}\end{center}
\vspace{-30pt}
\caption{Some of the products of links used in the Asqtad action.  The Naik term
is not shown.}
\label{asqtadlinks}
\end{figure}

%%The full set of fattened links is pictured below.
%%%The weights of each term are not shown.  
Each diagram in Fig.~\ref{asqtadlinks} represents a term
in the quark action of the form $\bar\chi(x) V(x,x+\hat\mu) \chi(x+\hat\mu)$,
where $V$ is the product of links along the path.

The ``fat-link'' contributions help to
suppress taste symmetry breaking.  The final
   5-link path was introduced by Lepage \cite{Lepage:1998vj} to correct the
   small momentum form factor.  The result is an action that has leading
errors of order $a^2 \alpha$ and  $a^4$.

%\begin{slide}{Systematic Errors}
The decision to base our calculations on the Asqtad action is based not
on a fondness for fractional powers, but on the expectation that 
we can get to the correct chiral physics more easily this way.  The chiral
extrapolation for the $u$ and $d$ quarks has been a challenge for LQCD
throughout its history.

To generate ensembles of gauge configurations,
we must select certain physical parameters: the
lattice spacing ($a$) or gauge coupling ($\beta$), 
the grid size ($N_s^3 \times N_t$), and the quark masses ($m_{u,d}$, $m_s$).

To control systematic errors, we must
take the continuum limit, 
take the infinite volume limit and
extrapolate to light quark mass for the $u$ and $d$ quarks.  However, we
can work at the physical $s$ quark mass.

%\begin{slide}{Ensemble of Configurations}

\begin{figure}[tbh]
\epsfxsize=0.99 \hsize
%\epsfysize=0.99 \hsize
%\vspace{-1.75in}
\epsfbox{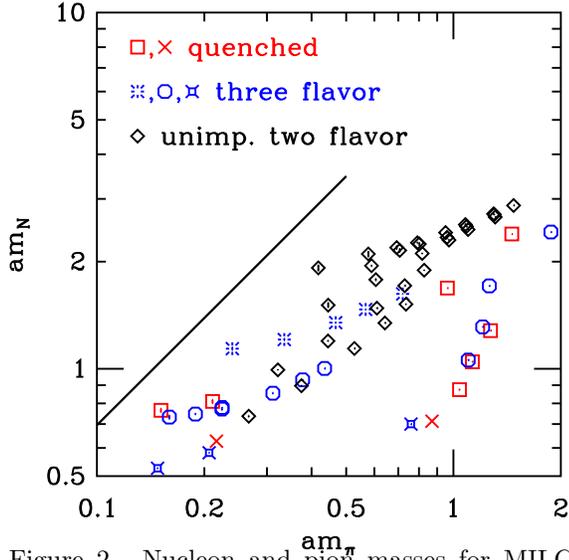}
\vspace{-36pt}
\caption{Nucleon and pion masses for MILC unimproved two-flavor
staggered runs %($\diamondsuit$) 
and improved runs with $N_f=0$ or 3.
%three dynamical quarks
%(burst, octagon, fancy plus), or in the quenched approximation (square, cross)
}
\label{mnucvsmpiplot}
\end{figure}

%%%For about four years MILC has been generating three flavor configurations to
%%%allow control of these errors.  The configurations are available to others
%%%through the NERSC Gauge Connection.  
In the Table \ref{paramtable} we give some of the details of the MILC ensembles.
We are able to go closer to the chiral limit than we did with
unimproved staggered quarks in the mid-90s.
This can be seen in Fig.~\ref{mnucvsmpiplot}, 
where the pion and nucleon masses are plotted
in lattice units.  The straight line corresponds to the physical
pion to nucleon mass ratio.  Approaching this line corresponds to the
chiral limit.  As the plot is a log-log plot, approaching the lower left corner
(and going beyond) corresponds to the continuum limit.  Note the three octagons
to the left of the lowest left-most diamond.  At this lattice spacing, with
2+1 flavors we have gone closer to the chiral limit.  One also sees fancy plus
symbols at smaller lattice spacing and comparable or smaller $m_\pi/m_N$.

\begin{table}[t]
\caption{Masses, coupling and number of configurations in MILC ensembles}
\vspace{4pt}
\label{paramtable}
\begin{center}
\begin{tabular}{|l|l|l|}
\hline
\multicolumn{3}{|c|}{
\rule[-4pt]{0pt}{14pt}
$a=0.12$ fm; $20^3\times64$ (coarse)}\\
\hline
\rule[-4pt]{0pt}{14pt}
$am_{u,d}$ / $am_s$  & \hspace{-1.0mm}$10/g^2$ & \# conf.\\
\hline
    0.40 /0.40   & 7.35 & 332 \\
    0.20 /0.20   & 7.15 & 341 \\
    0.10 /0.10   & 6.96 & 340 \\
    0.05 /0.05   & 6.85 & 425 \\
    0.04 /0.05   & 6.83 & 351 \\
0.03 /0.05   & 6.81 & 564 \\
0.02 /0.05   & 6.79 & 485 \\
0.01 /0.05   & 6.76 & 608 \\
0.007/0.05   & 6.76 & 447 \\
0.005/0.05   & 6.76 & 137 \\
\hline
\multicolumn{3}{|c|}{
\rule[-4pt]{0pt}{14pt}
$a=0.09$ fm; $28^3\times96$ (fine)}\\
\hline
\rule[-4pt]{0pt}{14pt}
$am_{u,d}$ / $am_s$  & \hspace{-1.0mm}$10/g^2$ & \# conf.\\
\hline
0.031  / 0.031   & 7.18 & 336 \\
0.0124  / 0.031   & 7.11 & 531 \\
0.0062  / 0.031   & 7.09 & 583 \\
%%%0.0031  / 0.031   & 7.08 & equil.*$40^3$ \\
\hline
\end{tabular}
\vspace{-24pt}
\end{center}
\end{table}

\section {RESULTS FOR HIGH PRECISION OBSERVABLES}

Certain quantities are intrinsically easier to calculate on the lattice
than others. For example, a ground
state mass is much easier to calculate than
an excited state mass in the same channel.  Also, a stable particle mass is 
much easier to calculate accurately than an unstable particle mass.
The Fermilab, HPQCD, MILC and UKQCD collaborations have been working together
to confront a variety of experimental results, and to make predictions
that can be tested.  %% (if we beat the experimenters).  
So far, we have 
mainly been concentrating on high precision observables.
If we fail to reproduce well known experimental results, after having
controlled the chiral, continuum and finite volume extrapolations, we
are left with little wiggle room.  That is, we can no longer blame
the quenched approximation.  We would either
conclude that staggered quarks are flawed, or that there is a problem
with (L)QCD.  On the other hand, if we are able to demonstrate good agreement
with a variety of measured quantities, it would give us confidence that
we are also able to make predictions, many of which are crucial for
accurate determination of standard model parameters such as quark masses
and CKM mixing matrix elements.

Our results depend on four input masses and the gauge coupling (equivalently
the lattice spacing).
The lattice spacing is determined from the $\Upsilon^\prime - \Upsilon$ 
mass difference, a quantity with little dependence on the $b$ quark mass 
\cite{Davies:1997mg}.
The common mass of the $u$ and $d$ quarks can be determined from $m_\pi$,
while $m_s$ is determined from $2 m_K^2-m_\pi^2$, or a
simultaneous fit to $m_\pi$ and $m_K$ determines $m_{u,d}$ and $m_s$.
The charm quark mass $m_c$ is determined from $m_{D_s}$, and $m_b$ is
determined from $m_\Upsilon$.

The results that follow are mainly the work of the following people
and collaborations \cite{Davies:2003ik}:

{C.~T.~H.~Davies}, {E.~Follana}, {A.~Gray},
{G.~P.~Lepage}, {Q.~Mason}, {M.~Nobes}, {J.~Shigemitsu},
{H.~D.~Trottier}, {M.~Wingate}:
{HPQCD and UKQCD Collaborations};

{C.~Aubin}, {C.~Bernard}, {T.~Burch}, {C.~DeTar},
{S.~G.}, {E.~B.~Gregory}, {U.~M.~Heller}, {J.~E.~Hetrick},
{J.~Osborn}, {R.~Sugar}, {D.~Toussaint}:
{MILC Collaboration};

{M.~Di Pierro}, {A.~El-Khadra}, {A.~S.~Kronfeld},
{P.~B.~Mackenzie}, {D.~Menscher}, {J.~Simone}:
{HPQCD and Fermilab Collaborations}.

The work on the topological susceptibility was done by MILC including
T.~DeGrand, A.~Hasenfratz and A. Hart \cite{Bernard:2003gq}.

\begin{figure}
\begin{center}
\begin{tabular}{c c}
\epsfxsize=0.47 \hsize
\epsfbox{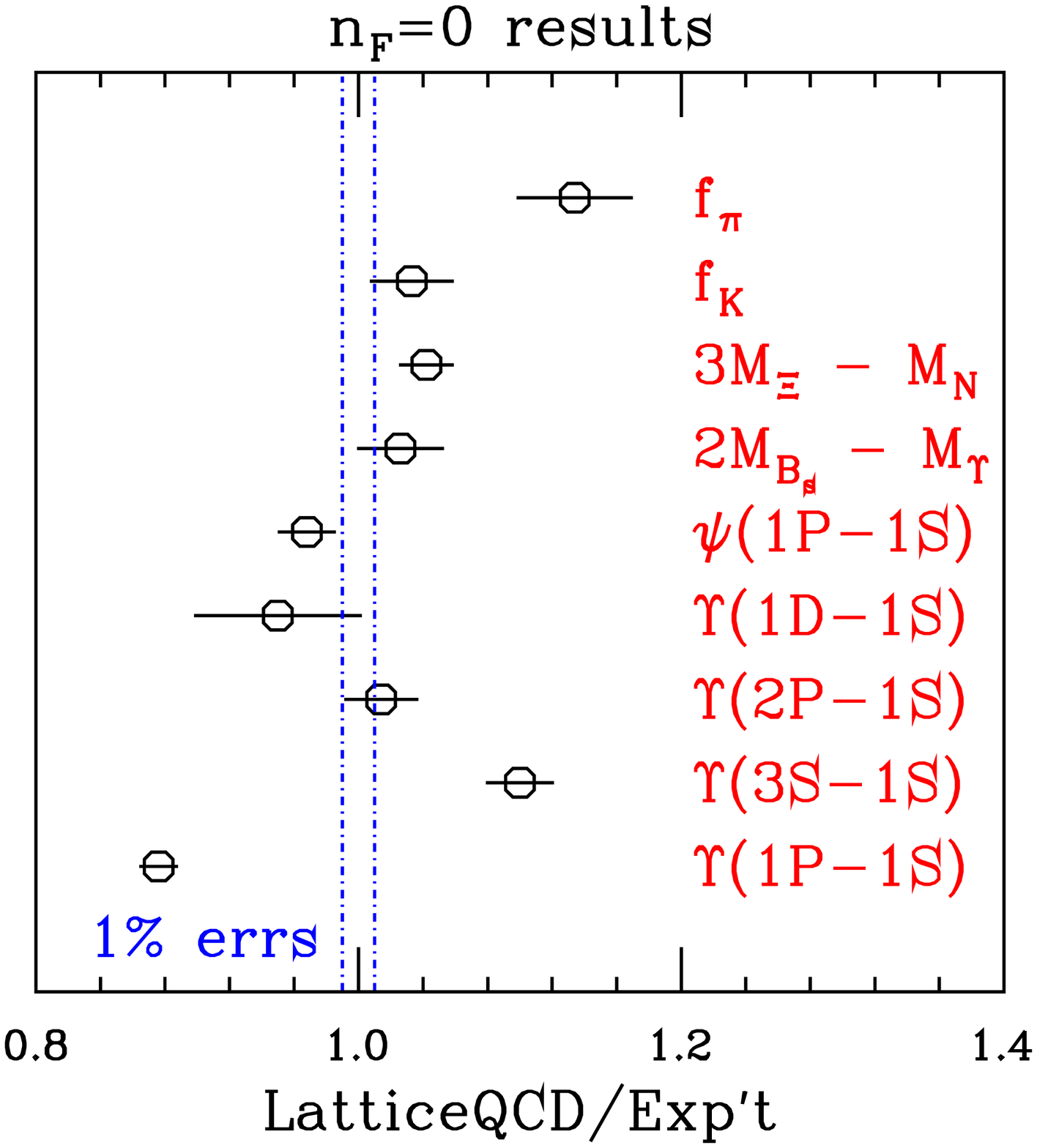}
&
\epsfxsize=0.47 \hsize
\epsfbox{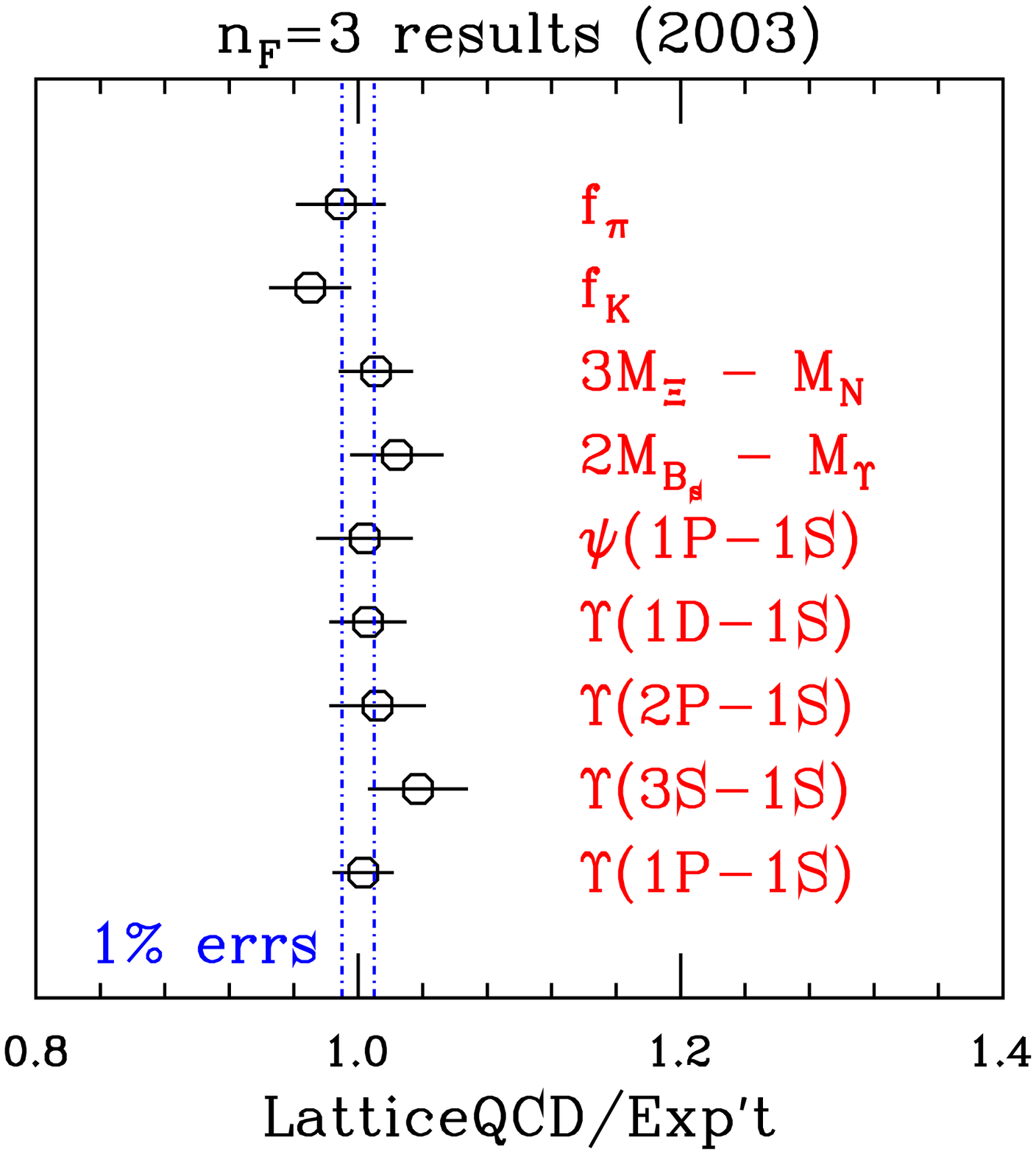}
\end{tabular}
\vspace{-12pt}
\caption{Ratio of lattice result to
experimental value for nine quantities.  Quenched (dynamical) results on the 
left (right).}
\label{ratioplot}
\end{center}
\end{figure}

\section {$\pi$, K MASSES AND DECAY CONSTANTS}
In our latest calculations, we have a greatly
improved understanding of chiral extrapolation due to five
factors: 1) inclusion of taste symmetry breaking,
2) improved action, 3) more and lighter quark masses, 4) partial quenching and
5) simultaneous fit to mass and decay constant results.

In MILC's prior work with two flavors of unimproved KS quarks,
we had five dynamical quark masses at several fixed values of $\beta$.
With no consideration for taste symmetry breaking, 
we tried seven combinations of powers and logs to try to fit the
pion masses, but our best fits had only a combined confidence level of 0.1.
In that work, the valence light quark mass was the same as the dynamical
mass, so there were only five data points for each $\beta$ value.
(In the current work with partial quenching, we have over 100 points for
each lattice spacing.)  

\begin{figure}[t]
\begin{center}
\epsfxsize= 0.99 \hsize
%\vspace{-.25in}
\epsfbox{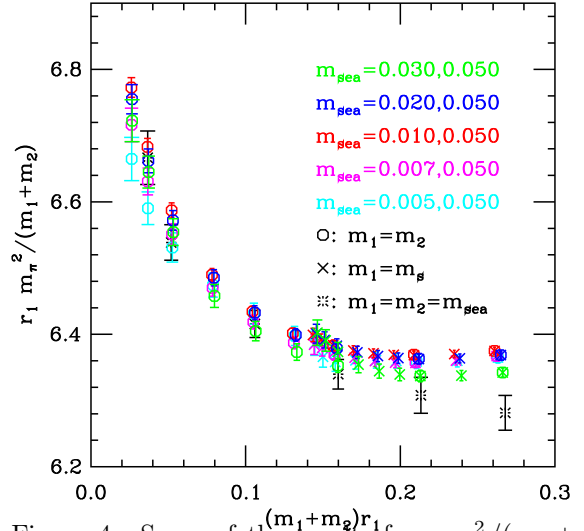}
\vspace{-36pt}
\caption{Some of the results for $r_1m_\pi^2/(m_1+m_2)$ on some of the coarse 
ensembles.  There are too many partially quenched results to plot all available
data.}
\label{coarsechiraldata}
\end{center}
\end{figure}

In Fig.~\ref{coarsechiraldata} we plot some of our
data for the five ensembles with the lightest  $m_{u,d}$ on the
coarse lattice \cite{Aubin:2003ne}.  Quark propagators for nine different valence quark masses
were calculated.   The octagons correspond to equal mass quark and antiquark.
The crosses have the antiquark fixed at the dynamical strange quark mass.
Many additional points with different combinations of the valence masses
are not plotted.  The bursts include results from ensembles with heavier
quarks than those listed in the legend.

%%\begin{slide}{Quenched Chiral Log Search 1996}
%%%\begin{center}
%%%\begin{figure}
%%%\epsfxsize= 0.99 \hsize
%%%%\vspace{-.25in}
%%%\epsfbox{../quenched_mpisq_over_mq_1996.ps}
%%%\caption{Compilation of results for $m_\pi^2/m_q$ for Lattice 96 review REF.
%%%Includes results from MILC, Kim and Sinclair, and Kim and Ohta.  DETAILS}
%%%\end{figure}
%%%\end{center}

%%%Can I find old confidence level of fits?

%%\begin{slide}{Asqtad dynamical results}

%\begin{slide}{Chiral fits (tasteless)}
%\vspace{-24pt}
The leading order formula from chiral perturbation theory for pion mass
and decay constant depends on four combinations of the Gasser-Leutwyler
constants.

\begin{eqnarray}
\lefteqn{\frac{M_{\pi}^2}{2\mu m_l} = 1 + 
\frac{1}{96\pi^2 f^2} [  3 M_{\pi}^2\log(M_{\pi}^2)
- M_\eta^2 \log(M_\eta^2) ]} \nonumber\\
 & \mbox{} + (2L_8-L_5)\frac{8}{f^2}M_{\pi}^2 \nonumber\\
 & \mbox{} + (2 L_6-L_4) \frac{12}{f^2} (M_{\pi}^2 + M_\eta^2) 
\end{eqnarray}
\BEA
f_{\pi}/f &= 1 + 
\frac{(-1)}{16\pi^2 f^2} \left[  M_{\pi}^2\log(M_{\pi}^2) \right. - \EL
&\left. \frac{1}{4} (M_\pi^2+M_\eta^2) \log ( \frac{1}{2}(M_\pi^2+M_\eta^2) )
 \right] \EL
&+ L_5\frac{4}{f^2}M_{\pi}^2 + L_4 \frac{6}{f^2} (M_{\pi}^2 + M_\eta^2)
\EEA

However, with staggered quarks there are taste symmetry violations that
modify these formulae to take into account the fact that all the pion states
are not degenerate with the Goldstone pion.
These formulae are quite complicated and may be found 
in Ref.~\cite{Aubin:2003mg}.
They involve new couplings $\delta'_V$ and $\delta'_A$ from the taste
symmetry breaking part of the effective Lagrangian.
%\begin{slide}{Chiral fits (with taste violation)}
%%Really too complicated to fit on a slide, see hep-lat/0304014 and
%%hep-lat/0306026.
%%\begin{eqnarray}
%%\frac{M_{\pi}^2}{2\mu m_l} 
 %%&=& \mu\Biggl\{1 + \frac{1}{16\pi^2 f^2}\Biggl(
         %%\biggl[ -4\ell(m^2_{\pi^0_V})
        %%- \frac{2a^2\delta'_V}{m_{\eta'_V}^2 - m^2_{\eta_V}}\biggl(
         %%\frac{m_{S_V}^2 - m^2_{\eta_V}}
        %%{m_{\pi^0_V}^2 -
        %%m^2_{\eta_V}} \ell(m^2_{\eta_V}) \nonumber\\*
        %%&&-  \frac{m_{S_V}^2 - m^2_{\eta'_V}}
        %%{m_{\pi^0_V}^2 -m^2_{\eta'_V}}  \ell(m^2_{\eta'_V})
        %%\biggr)\biggr]+\biggl[ V\to A \biggr]+
        %%\ell(m_{\pi^0_I}^2)
         %%- \frac{1}{3}\ell(m_{\eta_I}^2)\Biggr)\nonumber\\*
        %%&&+\frac{16\mu}{f^2}\left(2L_8-L_5\right)\left(2m_l \right)
        %%+\frac{32\mu}{f^2}\left(2L_6-L_4\right)
         %%\left(2m_l+m_s \right) + a^2 C \Biggr\}\nonumber
%%\end{eqnarray}
%%\begin{eqnarray}
 %%f_\pi && = f\Biggl\{ 1 +
        %%\frac{1}{16\pi^2 f^2}
         %%\Biggl[-\frac{1}{16}\sum_{B}\left(
          %%2\ell(m^2_{\pi^0_B})+ \ell(m^2_{K^+_B})\right)
         %%\nonumber \\* &&
        %%\!\!\!+ 2a^2\delta'_V\Biggl(  \frac{m^2_{S_V}-m^2_{\eta_V}}
        %%{(m^2_{\pi^0_V}-m^2_{\eta_V})(m^2_{\eta'_V}-m^2_{\eta_V})}
        %%\ell(m^2_{\eta_V})
        %%+ \frac{m^2_{S_V}-m^2_{\eta'_V}}
        %%{(m^2_{\pi^0_V}-m^2_{\eta'_V})(m^2_{\eta_V}-m^2_{\eta'_V})}
        %%\ell(m^2_{\eta'_V})\nonumber \\* &&
        %%+  \frac{m^2_{S_V}-m^2_{\pi^0_V}}
        %%{(m^2_{\eta_V}-m^2_{\pi^0_V})(m^2_{\eta'_V}-m^2_{\pi^0_V})}
        %%\ell(m^2_{\pi^0_V}) \Biggr)
        %%+ \Bigl( V\to A  \Bigr)\Biggr] \nonumber\\*&&
        %%+ \frac{16\mu}{f^2}\left( 2m_{l} + m_s\right)L_4
        %%+ \frac{16\mu}{f^2}m_{l} L_5
        %%+a^2 F \Biggr\} \ ,
%%\end{eqnarray}
%%where the sum over $B$ is over all 16 tastes of mesons, and where 
The formulae also involve the
chiral log function, $\ell$, given by 
\begin{equation}\label{eq:chiral_log}
        \ell( m^2) \equiv  m^2 \left(\ln \frac{m^2}{\Lambda_\chi^2}
         + \delta_1(mL)\right)\ ,
\end{equation}
with $\Lambda_\chi$ the chiral scale, $\delta_1$ the known finite volume
correction and $L$ the spatial size.

\begin{figure}
\epsfxsize= 0.99 \hsize
\epsfbox{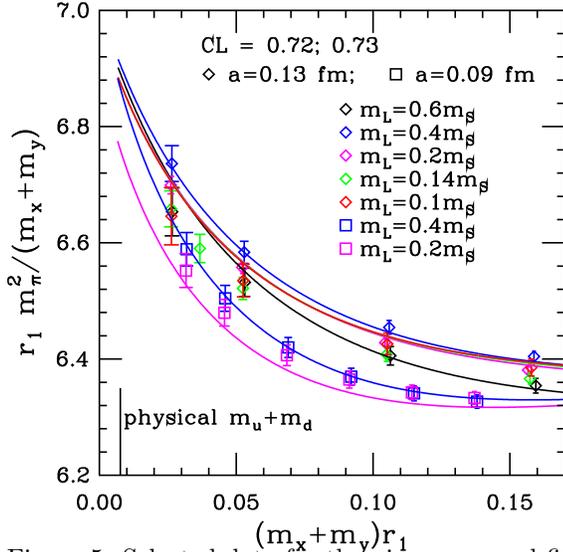}
\vspace{-36pt}
\caption{Selected data for the pion mass and fit curves from simultaneous
fit to mass and decay constant results.}
\label{pionchiralfit}
\end{figure}

\begin{figure}
\epsfxsize= 0.99 \hsize
\epsfbox{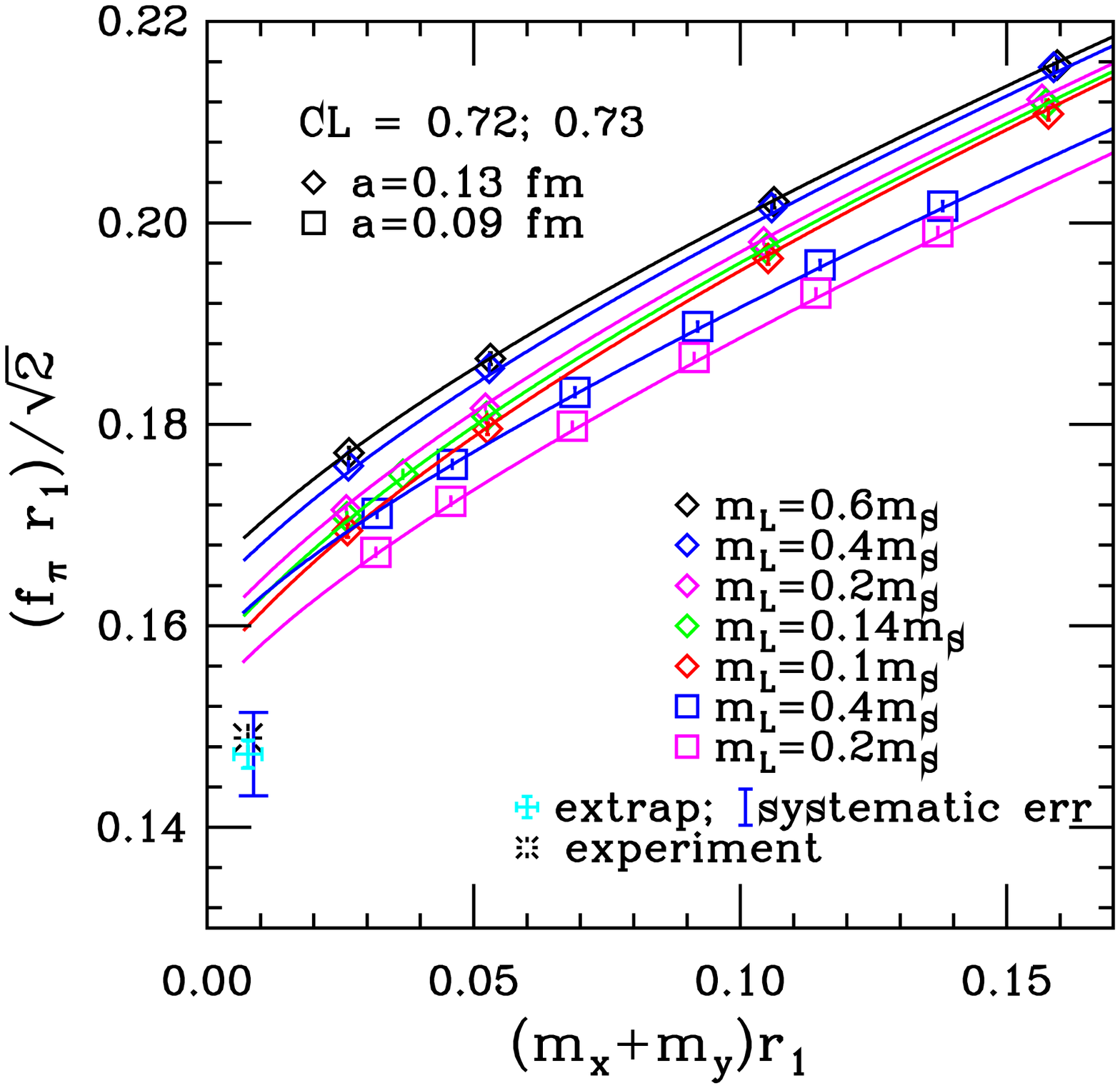}
\vspace{-36pt}
\caption{Selected data for pion decay constant and fit curves from simultaneous
fit to mass and decay constant results.}
\label{fpichiralfit}
\end{figure}

\begin{figure}[t]
\epsfxsize= 0.99 \hsize
\epsfbox{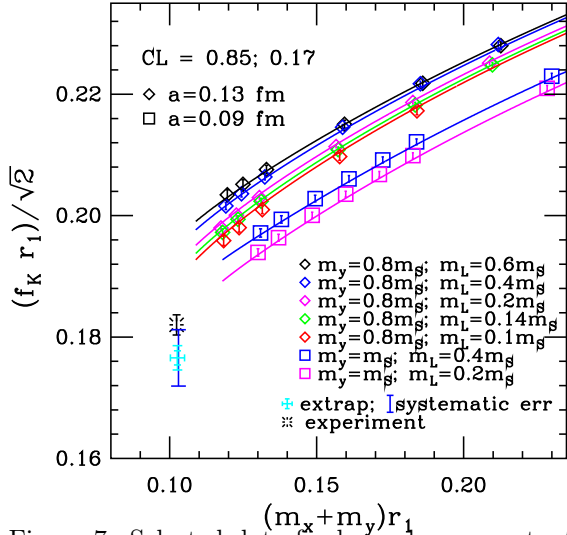}
\vspace{-36pt}
\caption{Selected data for kaon decay constant and fit curves from simultaneous
fit to mass and decay constant results.}
\label{fKchiralfit}
\end{figure}

In Figs.~\ref{pionchiralfit} and \ref{fpichiralfit}, 
we show results for $r_1 m_\pi^2/(m_x+m_y)$ and
$f_\pi r_1$ with $m_x$ and $m_y$ the two valence quark masses, and $r_1$
a scale set from the heavy-quark potential.  Only some of the data points
are plotted.  The curves correspond to partially quenched results on
ensembles with the light quark mass indicated in the legend.  Coarse (fine)
lattice results are plotted with diamonds (squares).  The confidence level of
the fit is 0.72, ignoring the contribution of the Bayesian constraints and
0.73 including them.  The points are plotted after taking into account
finite volume corrections that are sometimes as large as 1\%.
In Fig.~\ref{pionchiralfit}, a vertical bar shows the 
value of $m_u+m_d$, determined from
the physical pion mass.  The extrapolated value for $f_\pi$ and the 
experimental result are shown in the lower left of Fig.~\ref{fpichiralfit}.  
Note the close agreement.  
The systematic error includes variations in which points are included 
in the fit, variations in assumptions about how rapidly the 
taste-violating terms and non-taste violating terms are changing with
lattice spacing, and, most important, the scale error of 2.2\%.
In Fig.~\ref{fKchiralfit}, a similar plot is shown for $f_K$.  
The difference here is that more heavy valence mass points are
included in the $f_K$ fits.  One can see that the Bayesian constraints
contribute more to $\chi^2$ here.  Agreement with experiment is also
acceptable.  The dynamical strange quark mass used in these ensembles is
10--20\% higher than the result obtained for the valence $m_s$.  Additional
coarse ensembles are being generated with $am_s=0.03$.

\begin{figure}[t]
\begin{center}
\epsfxsize= 0.99 \hsize
\epsfbox{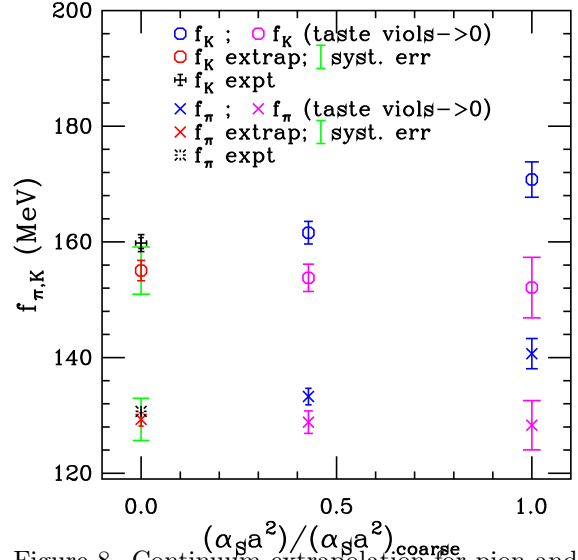}
\vspace{-36pt}
\caption{Continuum extrapolation for pion and kaon decay constants.}
\label{decayconstantcontinuumextrap}
\end{center}
\end{figure}

Figure~\ref{decayconstantcontinuumextrap} shows the continuum 
limit of $f_\pi$ and $f_K$. For these decay constants, the leading
correction is linear in $\alpha_s a^2$.  Values for the coarse and fine
results are also shown with taste symmetry violating terms set to zero.
When this is done, there is very little difference between the two
lattice spacings, supporting the contention that taste violations are
the major source of difference.

\subsection{Gasser-Leutwyler Parameters}
We now have preliminary results for several combinations of
Gasser-Leutwyler parameters using
$m_\eta$ as the chiral scale.  We find
\begin{eqnarray}\label{eq:Li_results}
% 2L_6 - L_4 &=& 0.48(15)({}^{+8}_{-26}) \; \times 10^{-3} \nonumber\\
% 2L_8 - L_5 &=& -0.13(6)({}^{+5}_{-28}) \; \times 10^{-3} \nonumber\\
% L_4 &=& 0.25(31)({}^{+66}_{-22}) \; \times 10^{-3} \nonumber\\
% L_5 &=& 1.92(34)({}^{+56}_{-17}) \; \times 10^{-3}\ , \nonumber
% round numbers:
2L_6 - L_4 &=& 0.5(2)({}^{+1}_{-3}) \; \times 10^{-3} \\
2L_8 - L_5 &=& -0.1(1)({}^{+1}_{-3}) \; \times 10^{-3} \\
L_4 &=& 0.3(3)({}^{+7}_{-2}) \; \times 10^{-3} \\
L_5 &=& 1.9(3)({}^{+6}_{-2}) \; \times 10^{-3}\ ,
\end{eqnarray}
where the first error is statistical and the second is systematic.
The systematic errors are dominated by differences over
various acceptable chiral fits.

In the continuum, it is expected that $L_5 = 2.3(2)\!\times\!10^{-3}$ and
$L_4 \approx L_6 \approx 0$ \cite{Donoghue:DSM}.
%%(from J. Donoghue, {\it et al}.  %%E. Golwich, and B. Holstein,
%%{\it Dynamics of the Standard Model}).  
%%(Cambridge University Press, New York, 1992), p. 166.)

The result for $2L_8-L_5$ of
$-0.1(1)({}^{+1}_{-3}) \; \times 10^{-3}$
is well outside the range that allows for
$m_u=0$ \cite{MUZERO}  %[Kaplan and Manohar; Cohen, Kaplan and Nelson] 
 which is (approximately)
\begin{equation}
-3.4\times 10^{-3} \le 2L_8-L_5 \le -1.8\times 10^{-3} \ . \nonumber
\end{equation}
Thus, if MILC's preliminary result holds up, it would rule out the
$m_u=0$ possibility.  This result seems likely to be robust under
changes in treatments of higher order terms.

\section{HEAVY QUARK SPECTRUM}
The most stringent tests of LGT come from hadrons that require no 
chiral extrapolation for the valence quarks,
are stable to strong decay, and far from threshold.
For example, J/$\Psi$, $\Upsilon$, $D_s$ and $B_s$ meet these conditions.
The HPQCD and UKQCD collaborations have used a non-relativistic QCD (NRQCD) 
\cite{Davies:1995db} 
action accurate to order $v^4$
to study bottomonium on MILC configurations.
The Fermilab collaboration has used clover quarks with the Fermilab
interpretation \cite{El-Khadra:1996mp} to study charmonium states.

\begin{center}
%\begin{sidewaysfigure}
\begin{figure}
\epsfxsize= 0.99\hsize
\epsfbox{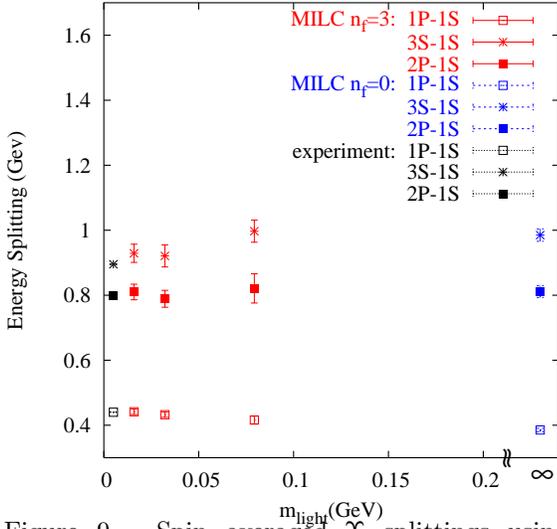}
\vspace{-36pt}
\caption{Spin averaged $\Upsilon$ splittings using NRQCD.}
\label{splitgnu}
\end{figure}
%\end{sidewaysfigure}
\end{center}

The spin averaged bottomonium splittings are shown in Fig.~\ref{splitgnu} 
for the 1P, 2P and 3S differences from the 1S level \cite{DAVIESWINGATE}.  
(Recall, the 2S-1S splitting 
is used to set the scale.)  Three coarse dynamical ensembles have been used.
The experimental results (without error bars)
are plotted near the left edge of the diagram.
On the right hand side of the diagram, the quenched values are plotted.
The corresponding values in Fig.~\ref{ratioplot} come from the quenched
results and lightest dynamical mass results shown here.

%\begin{slide}{Onium splitting w. NRQCD}

\begin{center}
\begin{figure}
\vspace{-36pt}
\epsfxsize= 0.99\hsize
\epsfbox{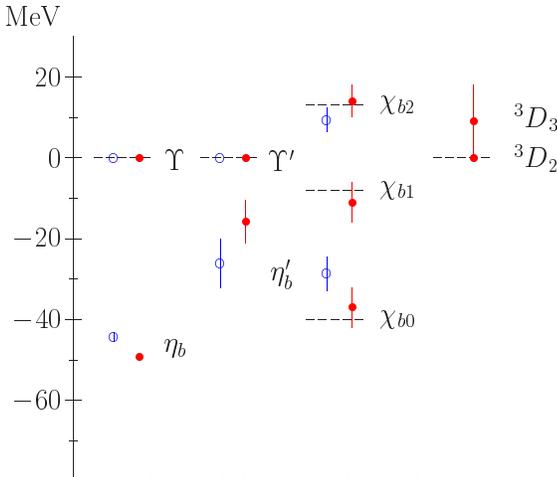}
\vspace{-36pt}
\caption{Several fine and hyperfine splittings in the $\Upsilon$ system
using NRQCD.  Experimental results indicated with dashed lines; open circles
denote quenched approximation, filled circles are from the coarse
ensemble with $am_{u,d}=0.01$, $m_s=0.05$.}
\label{upsilonfinestruc}
\end{figure}
\end{center}

%\begin{slide}{Fine structure}
Several fine and hyperfine splittings are shown in Fig.~\ref{upsilonfinestruc}
which contrasts results in the quenched approximation with results on one
dynamical ensemble with $m_{u,d}=0.2 m_s$.  There are significant
differences for the hyperfine splittings of $\Upsilon$ and $\Upsilon^\prime$.
However, the corresponding $\eta$ states have not been observed.
The fine structure in the $\chi$ states is improved on the dynamical
ensemble.

\begin{center}
\begin{figure}
\epsfxsize= 0.99\hsize
\epsfbox{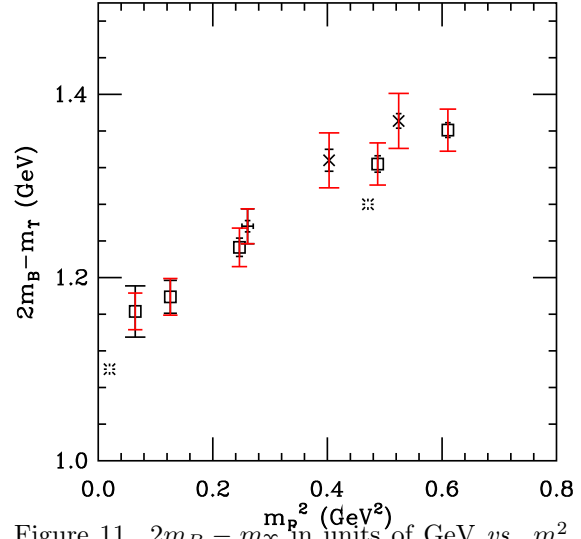}
\vspace{-36pt}
\caption{$2m_B-m_\Upsilon$ in units of GeV {\it vs}. $m_\pi^2$.  The 
experimental values for $B_{u,d}$ and $B_s$ are shown as bursts.}
\label{Del2MBU}
\end{figure}
\end{center}

Another quantity in the ratio plot is $2m_B-m_\Upsilon$.  Using NRQCD,
this quantity has been calculated on three of the ensembles ($m_{u,d}=0.01$,
0.02 and 0.03).  The splitting
is shown in Fig.~\ref{Del2MBU} for various valence quark masses.  
%The squares correspond
%to $m_{u,d}=0.2 m_s$.  There are also a few points from $m_{u,d}=0.4$ and 0.6
%$m_s$.  
The bursts are the experimental values.
For the ratio plot (Fig.~\ref{ratioplot}), a correction has been applied to
account for the fact that the dynamical strange quark mass is about 20\% too 
large.

\begin{figure}[t]
\begin{center}
\epsfxsize= 0.99\hsize
\epsfbox{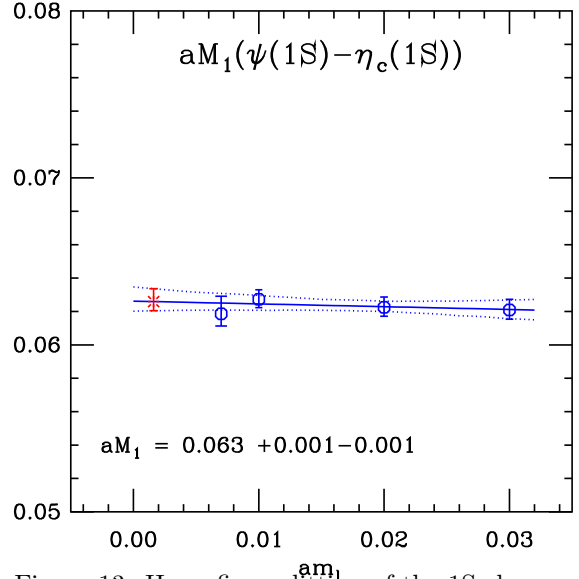}
\vspace{-36pt}
\caption{Hyperfine splitting of the 1S charmonium levels in lattice units
{\it vs}. the light dynamical mass in lattice units from four of the
coarse ensembles.}
\label{splitcharm}
\end{center}
\end{figure}

%\begin{slide}{Charmonium w. clover}
The Fermilab collaboration has looked at the spin averaged 1P--1S splitting
for charm quarks using clover/Fermilab quarks \cite{SIMONE}.  
Four dynamical ensembles
with $am_{u,d}=0.007$, 0.01, 0.02 and 0.03 have been analyzed.
Results are shown in Fig.~\ref{splitcharm}.
On the same ensembles, the hyperfine splitting between $\psi$ and $\eta_c$
has been calculated.  It is about 80\% of the experimental value.  This
splitting is sensitive to the coefficient of $\sigma_{\mu\nu}F_{\mu\nu}$.
This coefficient needs to be calculated to 1-loop level.  It is hoped
that using the corrected coefficient will improve the accuracy of this
result.

\begin{figure}[t]
\begin{center}
\epsfxsize= 0.99\hsize
\epsfbox{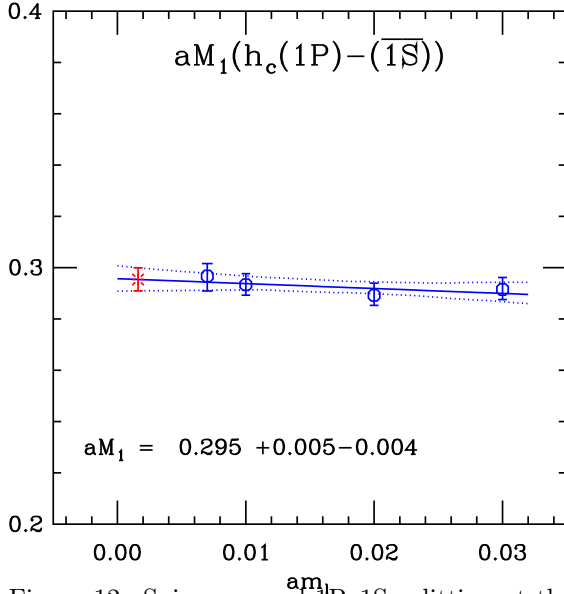}
\vspace{-36pt}
\caption{Spin averaged 1P--1S splitting at the charm mass in lattice
units {\it vs}. the light sea quark mass.}
\label{charmsplitting}
\end{center}
\end{figure}

%\begin{slide}{Charmonium hyperfine}

\section{LIGHT QUARK SPECTRUM}

\begin{figure}
\begin{center}
\epsfxsize= 0.99\hsize
\epsfbox{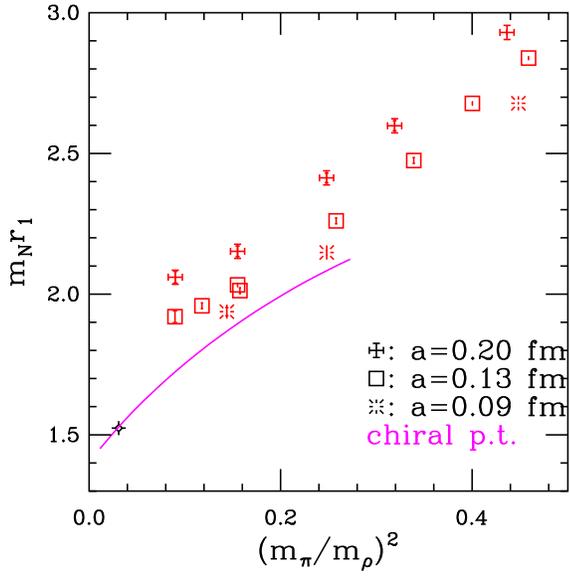}
\caption{Nucleon mass {\it vs}. $(m_\pi/m_\rho)^2$.}
\label{nucchiral}
\end{center}
\end{figure}

Hadrons containing up or down valence quarks require chiral extrapolation 
before they can be compared with experiment.  We have seen how this can be
done including taste symmetry breaking terms for the pseudoscalars.
For the nucleon, there is not yet a corresponding formula.
In Fig.~\ref{nucchiral}, we show $m_N r_1$ for the coarse and fine ensembles
as well as some ensembles with $a\approx 0.20$ fm.
The figure contains a phenomenological chiral perturbation theory curve, which
is not a fit to the data.  Although this figure may be suggestive that the
lattice results will approach the experimental result in the chiral and
continuum limit, more work is needed.  The nucleon is not yet a high precision
lattice quantity.

For the nucleon, we need to consider finite size effects carefully.
All our previous studies were done with two dynamical flavors and with
much larger taste symmetry breaking.  This case may have smaller
finite size effects.
For $m_{u,d}=0.2 m_s$ we are studying $28^3\times 64$ to complement
the $20^4\times 64$.  Preliminary results shows about a 1\% decrease
in mass on the larger volume.  More statistics are needed.
Another critical need is a chiral
fit to the nucleon masses including taste violations.

\begin{center}
\begin{figure}
\epsfxsize= 0.99\hsize
\epsfbox{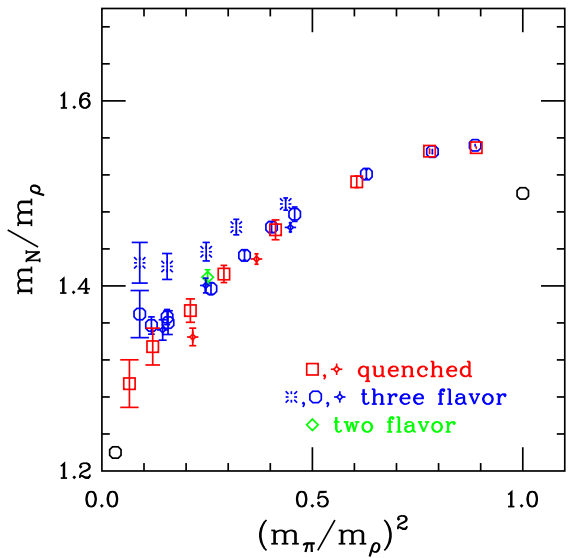}
\caption{APE plot for Asqtad improved KS quarks, with 0, 2 and 3 flavors.}
\label{MILCAPE03}
\end{figure}
\end{center}

With the additional {\it caveat} that the $\rho$ can decay and the lattice
fits take no particular account of this possibility, I present MILC's
current APE plot in Fig.~\ref{MILCAPE03}.

%\begin{slide}{APE plot}

\section{TOPOLOGY}
\begin{center}
\begin{figure}
\epsfxsize=3.0in
\epsfysize=3.0in
\vspace{-.25in}
\epsfxsize= 0.99\hsize
\epsfbox{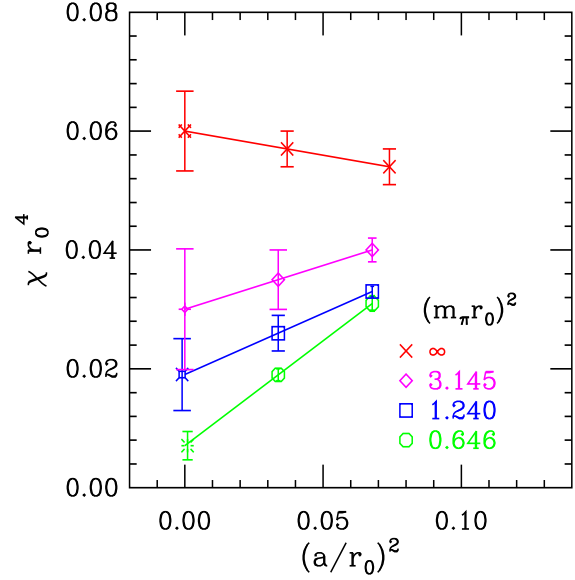}
\caption{Continuum limit of topological susceptibility.}
\label{continuumtopo}
\end{figure}
\end{center}

\begin{center}
\begin{figure}
\epsfxsize= 0.99\hsize
\epsfbox{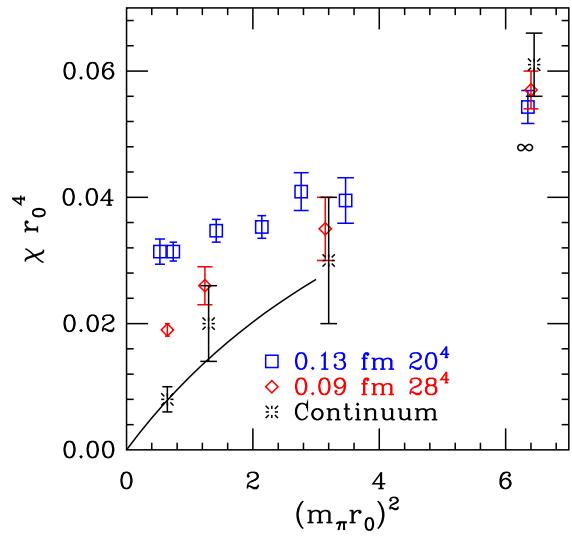}
\caption{Current topological susceptibility results on Asqtad ensembles.}
\label{topo2003}
\end{figure}
\end{center}

At Lattice 2002 there was a concern that the dynamical quarks were not
sufficiently suppressing topological susceptibility at light quark mass
\cite{Bernard:2002sa}.
%%%Figure~\ref{topo2002} shows those results and the leading order
%%%prediction for a theory with 2+1 flavors.  
Since then, the 
fine lattice runs have been extended \cite{Bernard:2003gq}.  
This has resulted in smaller errors,
and the 3-flavor point that was so high previously
%%%is so high in Fig.~\ref{topo2002} 
has come down.  
Figure~\ref{continuumtopo} shows an extrapolation to the
continuum limit in the quenched approximation and for the three fine lattice
ensembles.  We see in Fig.~\ref{topo2003} that the continuum extrapolations
are in reasonable agreement with the leading order theory.

%%%\begin{figure}
%%%\epsfxsize= 0.99\hsize
%%%\epsfbox{../chi_vs_mpi2_nf21.ps}
%%%\caption{Topological susceptibility results on Asqtad ensembles as of
%%%Lattice 2002.}
%%%\label{topo2002}
%%%\end{figure}
%%%\begin{figure}
%%%\epsfxsize= 0.99\hsize
%%%\epsfbox{../persistence.ps}
%%%\end{figure}

%%%\begin{center}
%%%\begin{figure}
%%%\epsfxsize= 0.99\hsize
%%%\epsfbox{../qb2896m031.ps}
%%%\end{figure}
%%%\end{center}

\section{PROSPECTS}

We feel that with the Asqtad action and 2+1 dynamical quarks,
we are making excellent progress.  
Most notably, the ratio plot is much improved compared with the quenched
approximation.  This is particularly true for the $b$-quark spectrum.
%Additional work not shown here on heavy-light decays and spectrum.
Much other work is in progress that could not be discussed in this
talk, and much additional work remains:
\begin{itemize}
\item Control chiral limit better on fine lattice
\item Study possible finite size effects further
\item Reduce lattice spacing to demonstrate control of continuum limit
\item Complete physics measurements on more ensembles, $f_B$, form factors, 
etc. \cite{MOREPHYS}
\item Use a highly improved FNAL-type heavy quarks \cite{OKTAY}
\item Develop new techniques for unstable particles and excited states
\item Complete additional perturbative calculations for 
renormalization and matching \cite{TROTTIER}
\end{itemize}

This work was supported by the U.S. Department of Energy under grant
FG02-91ER 40661.  I am pleased to thank these collaborators who provided
graphs or commented on the manuscript:  C.~Bernard
C.~Davies,
C.~DeTar,
A.~Gray,
U.~M.~Heller,
J.~Shigemitsu,
J.~Simone,
R.~Sugar,
D.~Toussaint,
W.~Wingate.


\begin{thebibliography}{99}
%\cite{Bernard:2001yj}
\bibitem{Bernard:2001yj}
C.~Bernard  [MILC Collaboration],
%``Chiral logs in the presence of staggered flavor symmetry breaking,''
Phys.\ Rev.\ D {\bf 65}, 054031 (2002)
[arXiv:hep-lat/0111051].
%%CITATION = HEP-LAT 0111051;%%

%\cite{Aubin:2002ss}
\bibitem{Aubin:2002ss}
C.~Aubin {\it et al.},
%``Chiral logs with staggered fermions,''
arXiv:hep-lat/0209066.
%%CITATION = HEP-LAT 0209066;%%

\bibitem{Blum:1996uf}
T.~Blum {\it et al.},
%``Improving flavor symmetry in the Kogut-Susskind hadron spectrum,''
Phys.\ Rev.\ D {\bf 55}, 1133 (1997)
[arXiv:hep-lat/9609036].
%%CITATION = HEP-LAT 9609036;%%
 
\bibitem{Lagae:1998pe}
J.~F.~Lagae and D.~K.~Sinclair,
%``Improved staggered quark actions with reduced flavour symmetry  violations for lattice QCD,''
Phys.\ Rev.\ D {\bf 59}, 014511 (1999)
[arXiv:hep-lat/9806014].
%%CITATION = HEP-LAT 9806014;%%
\bibitem{Lepage:1998vj}
G.~P.~Lepage,
%``Flavor-symmetry restoration and Symanzik improvement for staggered  quarks,''Phys.\ Rev.\ D {\bf 59}, 074502 (1999)
[arXiv:hep-lat/9809157].
%%CITATION = HEP-LAT 9809157;%%

%\cite{Orginos:1999cr}
\bibitem{Orginos:1999cr}
K.~Orginos, D.~Toussaint and R.~L.~Sugar  [MILC Collaboration],
%``Variants of fattening and flavor symmetry restoration,''
Phys.\ Rev.\ D {\bf 60}, 054503 (1999)
[arXiv:hep-lat/9903032].
%%CITATION = HEP-LAT 9903032;%%

%\cite{Bernard:1999xx}
\bibitem{Bernard:1999xx}
C.~W.~Bernard {\it et al.}  [MILC Collaboration],
%``Scaling tests of the improved Kogut-Susskind quark action,''
Phys.\ Rev.\ D {\bf 61}, 111502 (2000)
[arXiv:hep-lat/9912018].
%%CITATION = HEP-LAT 9912018;%%

%\cite{Davies:1997mg}
\bibitem{Davies:1997mg}
C.~T.~Davies, K.~Hornbostel, G.~P.~Lepage, P.~McCallum, J.~Shigemitsu and J.~H.~Sloan,
%``Further precise determinations of alpha(s) from lattice QCD,''
Phys.\ Rev.\ D {\bf 56}, 2755 (1997)
[arXiv:hep-lat/9703010].
%%CITATION = HEP-LAT 9703010;%%

%\cite{Davies:2003ik}
\bibitem{Davies:2003ik}
C.~T.~Davies {\it et al.}  [HPQCD Collaboration],
%``High-precision lattice QCD confronts experiment,''
arXiv:hep-lat/0304004.
%%CITATION = HEP-LAT 0304004;%%

%\cite{Bernard:2003gq}
\bibitem{Bernard:2003gq}
C.~Bernard {\it et al.},
%``Topological susceptibility with the improved Asqtad action,''
arXiv:hep-lat/0308019.
%%CITATION = HEP-LAT 0308019;%%

%\cite{Aubin:2003ne}
\bibitem{Aubin:2003ne}
C.~Aubin {\it et al.}  [MILC Collaboration],
%``Pion and kaon physics with improved staggered quarks,''
arXiv:hep-lat/0309088.
%%CITATION = HEP-LAT 0309088;%%

%\cite{Aubin:2003mg}
\bibitem{Aubin:2003mg}
C.~Aubin and C.~Bernard,
%``Pion and kaon masses in staggered chiral perturbation theory,''
Phys.\ Rev.\ D {\bf 68}, 034014 (2003)
[arXiv:hep-lat/0304014];
%%CITATION = HEP-LAT 0304014;%%
%\cite{Aubin:2003uc}
%\bibitem{Aubin:2003uc}
%C.~Aubin and C.~Bernard,
%``Pseudoscalar decay constants in staggered chiral perturbation theory,''
arXiv:hep-lat/0306026;
%%CITATION = HEP-LAT 0306026;%%
%\cite{Aubin:2003rg}
%\bibitem{Aubin:2003rg}
%C.~Aubin and C.~Bernard,
%``Staggered chiral perturbation theory,''
arXiv:hep-lat/0308036.
%%CITATION = HEP-LAT 0308036;%%

\bibitem{Donoghue:DSM}
J.~Donoghue, {\it et al}.  %%E. Golwich, and B. Holstein,
{\it Dynamics of the Standard Model},
(Cambridge University Press, New York, 1992), p. 166.

\bibitem{MUZERO}
D.~Kaplan and A.~Manohar,
Phys.\ Rev.\ Lett.\  {\bf 56} (1986) 2004;
A.~Cohen, D.~Kaplan and A.~Nelson,
JHEP {\bf 9911} (1999) 027.
%[arXiv:hep-lat/9909091].

%\cite{Davies:1995db}
\bibitem{Davies:1995db}
C.~T.~Davies, K.~Hornbostel, G.~P.~Lepage, A.~J.~Lidsey, J.~Shigemitsu and J.~H.~Sloan,
%``Precision charmonium spectroscopy from lattice QCD,''
Phys.\ Rev.\ D {\bf 52}, 6519 (1995)
[arXiv:hep-lat/9506026].
%%CITATION = HEP-LAT 9506026;%%

%\cite{El-Khadra:1996mp}
\bibitem{El-Khadra:1996mp}
A.~X.~El-Khadra, A.~S.~Kronfeld and P.~B.~Mackenzie,
%``Massive Fermions in Lattice Gauge Theory,''
Phys.\ Rev.\ D {\bf 55}, 3933 (1997)
[arXiv:hep-lat/9604004].
%%CITATION = HEP-LAT 9604004;%%

\bibitem{DAVIESWINGATE}
Thanks to A.~Gray, C.~Davies and M.~Wingate for providing these graphs from
the HPQCD/UKQCD collaborations.

\bibitem{SIMONE}
J.~Simone {\it et al}., these proceedings.

%\cite{Bernard:2002sa}
\bibitem{Bernard:2002sa}
C.~Bernard {\it et al.}  [MILC Collaboration],
%``Topological susceptibility with the improved Asqtad action,''
arXiv:hep-lat/0209050.
%%CITATION = HEP-LAT 0209050;%%
                                                                                
\bibitem{MOREPHYS}
C.~Bernard {\it et al}., these proceedings, arXiv:hep-lat/0309117; %Gregory
P.~Mackenzie {\it et al}., these proceedings;
M.~Okamoto {\it et al}., these proceedings, arXiv:hep-lat/0309107;
J.~Shigemitsu {\it et al}., these proceedings, arXiv:hep-lat/0309039;
M.~Wingate {\it et al}., these proceedings, arXiv:hep-lat/0309092.

\bibitem{OKTAY}
M.~Oktay {\it et al}., these proceedings.

\bibitem{TROTTIER}
H.~ Trottier, these proceedings; 
Q.~Mason, these proceedings;
M.~Nobes and H.~Trottier, these proceedings, arXiv:hep-lat/0309086.
\end{thebibliography}
\end{document}